\documentclass[prl,aps,twocolumn,groupedaddress,showpacs,floatfix,preprintnumbers,superscriptaddress]{revtex4-1}
\usepackage{amsmath,amssymb,multirow,bm}

\newcommand{\beq}{\begin{equation}}
\newcommand{\eeq}{\end{equation}}
\newcommand{\bea}{\begin{eqnarray}}
\newcommand{\eea}{\end{eqnarray}}
\usepackage{graphicx}
\usepackage[sort&compress]{natbib}
\usepackage{color}
\preprint{}

\begin{document}

%\begin{frontmatter}

\title{Relativistic effects in heavy ion Coulomb scattering}
\author{Ravinder Kumar}
\email{ravinderkumar12@yahoo.co.in}
\affiliation{Department of Physics and Astronomy, Texas A \& M University-Commerce, Commerce, Texas 75429, USA}
\affiliation{Department of Physics, Deenbandhu Chhotu Ram University of Science and Technology, Murthal 131039, Haryana, India}
\author{C.A. Bertulani}
\email{carlos.bertulani@tamuc.edu}
\affiliation{Department of Physics and Astronomy, Texas A \& M University-Commerce, Commerce, Texas 75429, USA}
\author{G. Robinson}
\email{grobinson5@leomail.tamuc.edu}
\affiliation{Department of Physics and Astronomy, Texas A \& M University-Commerce, Commerce, Texas 75429, USA}
\date{\today}

\begin{abstract}
The role of relativistic corrections in heavy ion Coulomb scattering at intermediate energies ($E_{lab}\gtrsim 50$ MeV/nucleon) is investigated by numerically solving a full set of coupled equations. We compare two methods: (a) one involving an exact account of interaction retardation with  (b) a method based on the expansion of effective Lagrangians in powers of the ion velocities, $v/c$.  Our study allows to infer the relevance of kinematic corrections, of retardation, and of magnetic interactions such as the Darwin force. We show that analytical formulas are able to describe all aspects of experimental interest of relativistic effects in heavy ion Coulomb scattering at intermediate energies without having to solve numerically the coupled equations.

\end{abstract}


\maketitle

\section{Introduction}

Properties of nuclei far from stability are not known at the level needed for an accurate description of several processes of interest for nuclear science. Therefore, much of the experimental effort in nuclear physics at present is dedicated to new radioactive beam facilities, the most expensive of them using secondary beams with high energy fragments obtained from primary collisions. By high energy here we mean energies of the order of 50 MeV/nucleon and above such as those in use at RIKEN/Japan, GANIL/France, GSI/Germany and NSCL/USA. New facilities are under construction, e.g., the FAIR facility in Germany and the FRIB facility in the USA. High energy radioactive beams  have fostered  the use of indirect techniques using reactions of rare nuclear isotopes with  the purpose of studying the structure of exotic nuclei  \cite{BERTULANI1993281,0034-4885-77-10-106901}  and nuclear astrophysics \cite{Bertulani2010195,Bertulani201656}.
 
Coulomb excitation is one of the main indirect techniques used in radioactive beam facilities mainly because the Coulomb interaction is well understood and also because it is intimately related to processes involving real photons like photo-absorption and gamma-decay of interest for studying nuclear structure and many processes of astrophysical interest \cite{0034-4885-77-10-106901}. Recent experiments  with Coulomb excitation have been used to unravel the physics of pigmy dipole resonances, dipole polarizability, energy density functionals, neutron skins,  equation of state of nuclear matter, etc \cite{Aumann2005,PhysRevLett.95.132501,PhysRevLett.102.092502,PhysRevLett.107.062502,PhysRevC.85.041304,1402-4896-2013-T152-014012,PhysRevC.92.031305,ALTSTADT2014197,Krumbholz20157,Marganiec2016200}. Experimental analyses assume that Coulomb scattering dominates the reaction process at forward angles, which is supported by theory for the scattering of heavy ions and of light nuclei with small binding energies \cite{BERTULANI1993281,0034-4885-77-10-106901}. In particular, elastic scattering of heavy ions is dominated by the Coulomb interaction up to the rainbow angle which reflects the onset of the nuclear interaction \cite{Bertulani1999139}. Since the analysis of Coulomb excitation experiments is based on the same premises, and since such reactions are carried out with kinetic energies consisting of a sizable fraction of the projectile's rest mass, it is imperative to account for relativistic effects not only in the kinematics (which is usually done in the experimental analysis),  but also in the reaction dynamics. This has often been overlooked both in theory and in experiments, except for a few theoretical studies \cite{Matzdorf1987,PhysRevC.42.2180}. It is the goal of this work to make a detailed assessment of this problem and to propose best ways to account for relativistic effects in Coulomb scattering of nuclei at intermediate and high energies collisions ($E_{lab} \gtrsim 50$ MeV/nucleon).

At low energies when the velocity of the projectile is much smaller than the speed of light, $v\ll c$, heavy ion collisions are well described by Rutherford scattering formulas except for minor corrections caused by Coulomb excitation, electron screening, or vacuum polarization. However, at intermediate and high energies when the speed of the projectile is comparable to the speed of light, relativistic effects  play a significant role. Therefore, an accurate knowledge of elastic Coulomb scattering at intermediate and high energy collisions is of great relevance for calibration of nuclear reaction experiments and to extract excitation amplitudes induced by the Coulomb interaction. Coulomb excitation at intermediate and high energy collisions of heavy ions is a very important tool in experimental nuclear physics and experimental analyses depend on a good understanding of dynamical relativistic effects  \cite{Bertulani1999139}. 

An early work on the effects of retardation in Coulomb scattering has been carried out by Matzdorf et al. \cite{Matzdorf1987} using classical trajectories which are well justified for heavy ion collisions. Another publication by Aguiar et al.  \cite{PhysRevC.42.2180} tackled the same problem using a perturbation expansion of the relativistic Lagrangian for the two-body Coulomb scattering. In Ref. \cite{Matzdorf1987}  retardation effects on the trajectory of one particle upon another via their mutual time-dependent electromagnetic fields were accounted for in a covariant way, accompanied by simplifying approximations to make the problem more manageable. They have investigated deflection angles, differential cross sections and the deviations of the time-dependent trajectory from non-relativistic Rutherford scattering. They also reported that the action of mutual magnetic fields  are rather small in the velocity range from 0.1c to 0.99c. This was shown specifically for Xe + U reaction. However, the relativistic mass correction effect was reported to be quite significant. Analytic formulas for light projectiles colliding with heavy targets have been obtained which are quite useful for a quick estimate of relativistic corrections in elastic differential cross sections. We show that such formulas also work exceptionally well for more symmetric systems.

Aguiar et al. \cite{PhysRevC.42.2180} have also studied relativistic effects in Coulomb scattering at intermediate energies. They have used an effective theory approach based on the  expansion of the classical electromagnetic Lagrangian in powers of $v/c$, where $v$ is the projectile velocity and $c$ the speed of light. They have considered corrections of the classical Lagrangian up to order $(v/c)^2$. But for particles with equal charge to mass ratio they have extended the formalism to include corrections up to $(v/c)^4$.  Analytical formulas have also been proposed to estimate the relativistic corrections and their contribution to differential cross sections. 

 In view of the large experimental interest of reactions in radioactive beam facilities, and the relevance of the Coulomb interaction for experiments, we have studied in  this work the relativistic effects in Coulomb scattering of nuclei. Much of our analysis is based on a comparison of the two different approaches to relativistic corrections in Coulomb scattering presented in Refs. \cite{Matzdorf1987} and \cite{PhysRevC.42.2180} for elastic collisions at intermediate and high beam energies. We have made qualitative and quantitative predictions for reactions with symmetric and asymmetric systems. Most importantly, we have shown that the problem is treatable with basic analytical methods.  In the next section we present a summary of the theoretical methods involving a full account of retardation and another using effective Lagrangians. 
  
\section{Theoretical Formalism}
\subsection{Covariant formulation}
In the following, the target is assumed to be located at the center of the coordinate system and the projectile is assumed to move towards the target parallel to the x-axis. The covariant equation of motion for a charged particle moving in an external electromagnetic field of another charged particle is given by \cite{jackson_classical_1999}
\begin{equation}
\frac{dp^{\alpha}}{d\tau} = \frac{q}{c}F^{{\alpha}{\beta}} U_{\beta},
\end{equation}
where $ p^{\alpha} $ and $ U_{\beta} $ are the 4-momentum and the 4-velocity respectively. $ q $ is the charge of one of the particles, $ \tau $ is the proper time of the considered particle and $ F^{\alpha\beta} $ is field strength tensor, which can be written in terms of the components  of the electric and magnetic fields $ {\bf E}$ and ${\bf B} $ due to the other particle.
To solve this equation of motion for the two-body scattering it is assumed that a projectile with charge $ q_p $ moves in the external field generated by target and vice versa. In terms of electric, $ \textbf{E} ^{(t)}$, and magnetic, $ \textbf{B} ^{(t)}$, fields of the target acting on the projectile, the coupled set of equations for motion for the projectile can be written as \cite{jackson_classical_1999}.
\begin{eqnarray}
&& \gamma^4(\textbf{u} \cdot \dot{\textbf{u}})= \gamma \frac{q_p}{m_p c} {\bf E}^{(t)}\cdot {\bf u}\nonumber
\\
&& \dot{{\bf u}}\gamma^2+\gamma^4 {\bf u} (\textbf{u} \cdot \dot{\textbf{u}})= \gamma \frac{q_p}{m_p c} \left({\bf E}^{(t)} +{\bf u} \times {\bf B}^{(t)}\right)
 \label{udot1}
\end{eqnarray}
where $ \textbf{u}=(u_1, u_2, u_3)={\dot{\textbf{x}}}/{c} $ is the  projectile velocity, $ {\bf x}$ is its position, $ m_p $  its rest mass, and $\gamma=1/{\sqrt{1-\beta^2}}$ is the Lorentz factor, with $\boldsymbol{\beta}={\bf u}/c$ representing the projectile velocity in units of $c$. The electric and magnetic fields are calculated at the retarded time $T$ defined in the text after Eq. \eqref{udot2} below. Since the motion is restricted to a scattering plane, only two of the coordinates, e.g., $x$ and $y$, need to be considered. 

In Ref. \cite{Matzdorf1987} it was shown that the above equations, together with similar equations for the target motion, lead to a full set of coupled equations for the motion of the projectile and the target. Numerically, one first looks at the effect of the retarded ${\bf E}^{(t)}$ and ${\bf B}^{(t)}$ fields generated by the target at the position of the projectile and then, after a time step, one corrects for the position of the target by reversing the roles of the target and the projectile in the equations above. This procedure is repeated from the initial position of the system until the effects of the fields at large distances become negligible. It was also shown  that if one neglects the magnetic field in the equations above  one obtains a much simpler set of coupled  equations for the projectile and target motion in the $x-y$ plane. A full and detailed derivation of these equations  are provided in Ref. \cite{Matzdorf1987}, where it was also shown that the inclusion of the magnetic field ${\bf B}$ amounts to a less than 1\% change for the scattering deflection angle and cross sections. 

For the projectile motion, these simplifications lead to the equations of motion
\begin{eqnarray}
 \dot{u_{1}}&=&\frac{q_p q_t}{m_{0}^{(p)} \gamma^3}    \frac{(\gamma^{-2}+u_2^2)n_1-u_1u_2n_2}{R^2[(\gamma^{-2}+u_1^2)(\gamma^{-2}+u_2^2)-u_1^2u_2^2]}
\nonumber \\
\dot{u_{2}}&=&\frac{q_p q_t}{m_{0}^{(p)} \gamma^3}    \frac{(\gamma^{-2}+u_1^2)n_2-u_1u_2n_1}{R^2[(\gamma^{-2}+u_1^2)(\gamma^{-2}+u_2^2)-u_1^2 u_2^2]}
\label{udot2}
\end{eqnarray}
where  ${\bf R}={\bf x}-{\bf r}(T)$ is the radius vector of the projectile location with respect to the target at position ${\bf r}(T)$ at the retarded time $T$, satisfying the retardation condition $(t-T)-R/c=0$. R is magnitude of the radius vector and $n_1$, $n_2$ are \textit{x} and \textit{y} components of the unit vector $\textbf{n}$ along the ${\bf R}$ direction. A similar set of equations as in \eqref{udot2} is solved for the target motion with the roles of the projectile and target reversed. This yields 4 coupled equations to be solved simultaneously. The Lienard-Wiechert  acceleration terms are not included in these equations because the modifications of Coulomb trajectories in heavy ion  collisions due to the  emission of radiation are extremely small. It is worthwhile noticing that for $u \ll 1$ and $\gamma \rightarrow 1$, these equations reduce to the well-known non-relativistic equations for the motion of a charged particle in the electromagnetic field generated by another charged particle.

Numerically the scattering angle is obtained as follows
\begin{equation}
\Theta({\ t\rightarrow +\infty})= \arctan\frac{dy}{dx}(t) \label{theta}
\end{equation}
by starting monitoring the scattering at a very large negative time for a collision with impact parameter $b$. Repeating the procedure for several impact parameters, the differential scattering cross section can be calculated from 
\begin{equation}
\frac{d\sigma}{d\Omega}=\frac{b(\Theta)}{\sin\Theta}  \left|\frac{db}{d\Theta}\right|. \label{dsigma}
\end{equation}

A simplified analytical formula was presented in Ref. \cite{Matzdorf1987}, valid when one collision partner remains nearly at rest, i.e. when the mass of the projectile is much smaller than the mass of target. In this case, the analytical approximations for the scattering angle and the differential cross section are given by   
\begin{equation}
\Theta= \pi-\frac{2 {\rm arccot}(k)}{\sqrt{1-k^2(b)\beta^2}},\label{theta1}
\end{equation}
where here $\beta=v_\infty/c$, and
\begin{equation}
\frac{d\sigma}{d\Omega}=\frac{b^2}{\sin\Theta} \left|\frac{(1+k^2(b))\xi^2}{2(1+k^2(b))k^2(b)\beta^2 \left(\displaystyle{
\frac{\pi-\Theta}{2}
}\right)-2\xi k(b) }\right|, \label{dsdo}
\end{equation}
with $ k(b)= ({d}/{2b})\sqrt{1-\beta^2} $,  $ d={2 q_pq_t}/({m_{p} v_\infty^2}) $ and $\xi(b)=\sqrt{1-k^2(b)\beta^2}$.  

Contrary to what was stated in Ref. \cite{Matzdorf1987}, we will show that these equations reproduce with high precision the numerical results obtained with Eqs. \eqref{udot2} even for symmetric systems, i.e. when the masses of the particles are comparable. This is achieved by replacing the projectile mass in the definition of the variable \textit{d} by the reduced mass of the system. There is no ab-initio justification for this step, except that we know that Eq. \eqref{udot1} reduces to the usual Coulomb scattering when $\gamma \rightarrow 1$ and $u \rightarrow 0$, as can be readily verified. Solving these equations numerically for the projectile motion and for the target motion simultaneously yields the practical net result of a one-body motion with a reduced mass, as is well known in non-relativistic classical mechanics.
 
\subsection{Effective Lagrangian method}
Ref.   \cite{PhysRevC.42.2180} has also studied the influence of relativistic corrections in Coulomb scattering at intermediate and high energies by means of  an expansion of the classical Langrangian to leading-oder (LO), next-to-leading order (NLO) and next-to-next-to-leading-order (NNLO) in powers of $v/c$, ${\cal L} = {\cal L}^{(LO)} + {\cal L}^{(NLO)}+{\cal L}^{(NNLO)}$ with
\begin{eqnarray}
&&{\cal L}^{(LO)}={1\over 2} \mu v^2 - {q_tq_p \over r},\nonumber \\
&&{\cal L}^{(NLO)}={\mu^4\over 8c^2}\left[ {1\over m_p^3} +{1 \over m_t^3}\right] v^4-{\mu^2 q_tq_p\over  2m_pm_t c^2r}  (v^2+v_r^2),\nonumber \\
&&{\cal L}^{(NNLO)}={m v^6 \over 512c^2}+{q_tq_p\over 16c^2r}\left[{1\over 8}(v^4-3v_r^4+2v_r^2v^2)\right. \nonumber\\ 
&&+\left. {q_tq_p\over m r}(3v_r^2-v^2) +{4 q_t^{2}q_p^{2}\over  m^2r^2}  \right], \label{lag}
\end{eqnarray}
with $\mu$ equal to the reduced mass, $v_r={\bf v}\cdot {\bf r}/r$ and {\bf v}(t) is relative velocity. The ${\cal L}^{(NNLO)}$ Lagrangian is only valid for symmetric systems with $m_p=m_t=m$.

The NLO Lagrangian is obtained by neglecting radiation and assuming instantaneous interactions between the particles \cite{jackson_classical_1999}. The first term accounts for the increase of masses due to relativity and the second term arises from the magnetic interaction between the particles. It is known as the Darwin interaction. When particles have the same charge to mass ratio, as in the case of identical particles, the dipole radiation vanishes and it is possible to derive the above NNLO Lagrangian from the Lagrangian of the two-particle classical electrodynamics  \cite{jackson_classical_1999}.  The first term is again another correction to the relativistic mass and the following one is due the corrections to the Darwing interaction. 

Inserting Eqs. \eqref{lag} into Euler-Lagrangian equations one gets a set of coupled equations for the relative position and momentum (velocity) of the particles as a function of time as discussed in detail in Ref.\cite{PhysRevC.42.2180}. Numerically the scattering angle and differential cross sections are determined by making use of Eqs. \eqref{theta} and \eqref{dsigma}. It is worthwhile mentioning that the modifications of Coulomb trajectories of heavy ion collisions due to the emission of radiation  are extremely  small \cite{BALANTEKIN1997324}. This justifies the use of both methods employed in Refs.  \cite{Matzdorf1987,PhysRevC.42.2180} without the inclusion of radiation.
 
Ref.   \cite{PhysRevC.42.2180} has also proposed analytic formulae when the mass of the projectile is much smaller than the mass of the target i.e. $m_p \ll m_t$. They obtained an analytical formula for the scattering angle given by the same equation as Eq. \eqref{theta1}. Their analytical approximation for the differential cross section is given by 
\begin{eqnarray}
\frac{d\sigma(v/c,\Theta)}{d\Omega}&=&\left[\frac{q_pq_t}{2\mu v^2\sin^2(\Theta/2)}\right]^2\left[1+g(\Theta)\frac{v^2}{c^2}\right. \nonumber \\ 
&+&\left. {\cal O}\left(\frac{v^4}{c^4}\right)\right]\label{thea2}
\end{eqnarray}
where here $v\equiv v_\infty$ is a short notation for the projectile velocity in the laboratory system at large distances,  $\mu$ is reduced mass of the system  and 
$g(\Theta)=3-\left[2+\left\{1+(\pi-\Theta)\cot \Theta\right\}\tan^2({\Theta}/{2})\right] $.\\

\section{Results and Discussion} 
The coupled equations of motion, Eqs. \eqref{udot2}, have been solved numerically by using an adaptive stepsize control Runge-Kutta method \cite{Press:1993:NRF:563041}. As  initial condition it is assumed that the target is at rest at the origin of the coodinate system and at time $t=-\infty$  the impact parameter is $b ({\ t\rightarrow-\infty})=y(t)$ with the projectile moving towards the target along the \textit{x}-axis with velocity $v_\infty$. As the projectile approaches the target the Coulomb interaction deflects it to a scattering angle at time $t=+\infty$. 
Through out the calculations the total trajectory length is kept around 80,000 fm to account for the long range of the Coulomb interaction. The calculation is repeated for several impact parameters $b$ varied from the sum of the nuclear radii $R_P+R_T$, with $R_i=1.2A_i^{1/3}$ fm, to $100$ fm in very small, $\Delta b =0.1$ fm, interval  steps. The precision of the computed differential cross section using Eqs. \eqref{theta} and \eqref{dsigma} is checked by comparison with the well-known non relativistic domain, the Rutherford differential cross section. In each case the relative error was found to be less than 1 part in $10^4$. 

\begin{figure}[ptb]
\begin{center}
\includegraphics[
width=3.4in]{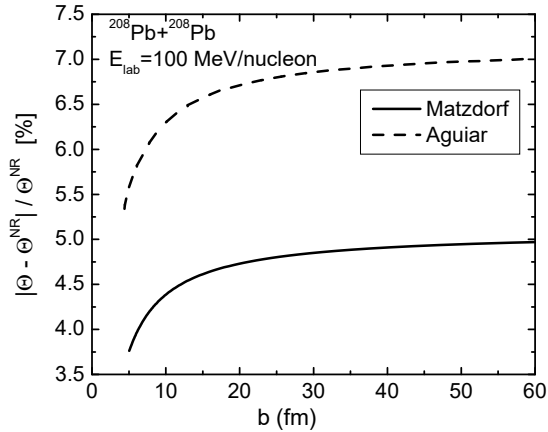}
\end{center}
\caption{Absolute value of the relative difference (in percent) between the methods of Matzdorf et al. \cite{Matzdorf1987} and of Aguiar et al. \cite{PhysRevC.42.2180} with the non-relativistic scattering angle $\Theta^{NR}=2\arctan{(q_pq_t/\mu v_\infty^2 b)}$  for $^{208}$Pb + $^{208}$Pb collisions at the laboratory energy of 100 MeV/nucleon. The dashed line is a numerical caclulation following the method of  Aguiar et al. \cite{PhysRevC.42.2180}  considering relativistic corrections up to order $(v/c)^{4}$ and the solid line is a numerical calculation for the  corresponding method of Matzdorf et al. \cite{Matzdorf1987}. The horizontal axis represents the impact parameter $b$ (in fm).}
\label{fig1}
\end{figure}

In Figure \ref{fig1} we plot the relative difference (in percent) between the numerical calculations following the methods of Matzdorf et al. \cite{Matzdorf1987} and of Aguiar et al. \cite{PhysRevC.42.2180} with the non-relativistic scattering angle $\Theta^{NR}=2\arctan{(q_pq_t/\mu v_\infty^2 b)}$  for $^{208}$Pb + $^{208}$Pb collisions at the laboratory energy of 100 MeV/nucleon. The dashed line is a numerical calculation following the method of  Aguiar et al. \cite{PhysRevC.42.2180} considering relativistic corrections up to order $(v/c)^{4}$ and the solid line is the numerical result for the  corresponding model of  Matzdorf et al. \cite{Matzdorf1987}. The horizontal axis represents the impact parameter $b$ (in fm). We observe that the method adopted by Matzdorf et al. yields a reduced correction for the non-relativistic scattering angle as compared to the method adopted by Aguiar et al. Since magnetic interactions are known to be small, the difference can be ascribed to the correct account of retardation implicit in the method adopted by Ref. \cite{Matzdorf1987}. It is also worthwhile noticing that the deviation from the classical Rutherford  scattering angle is smaller at smaller impact parameters, though not negligible either. The relativistic corrections increase and reach a nearly constant value of  $\sim 6.5 - 7$\% at larger impact parameters, i.e. at very forward scattering.

\begin{figure}[ptb]
\begin{center}
\includegraphics[
width=3.4in]{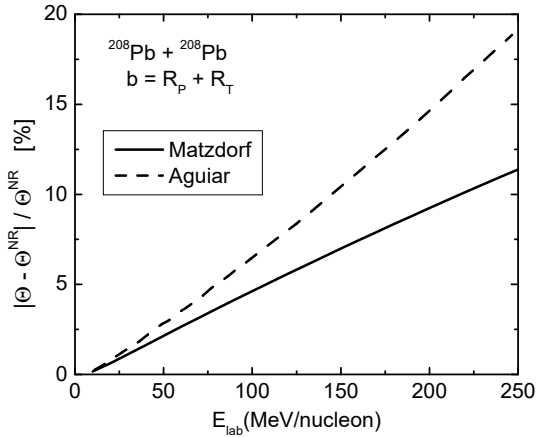}
\end{center}
\caption{Same as Fig. \ref{fig1}, but for a collision at grazing impact parameter $b=R_{P}+R_{T}$, as a function of the laboratory energy $E_{lab}$ (in MeV/nucleon).}
\label{fig2}
\end{figure}

The deviations from the classical Rutherford scattering increase with the bombarding energy, as expected. This is shown explicitly in  Fig. \ref{fig2} for a collision at grazing impact parameter $b=R_{P}+R_{T}$, as a function of the laboratory energy $E_{lab}$ (in MeV/nucleon). Not only the relativistic corrections become more important as the energy increases, but the effects of retardation also modify these corrections appreciably. The consideration of the relativistic mass increase without a corresponding account of retardation, overshoots the corrections due to relativity, as displayed by the dashed line obtained with the method of Ref.  \cite{PhysRevC.42.2180}.

\begin{figure}[ptb]
\begin{center}
\includegraphics[
width=3.4in]{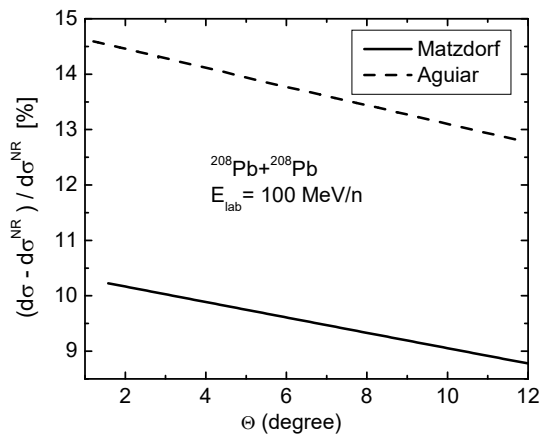}
\end{center}
\caption{Relative difference (in percent) between the methods of Matzdorf et al. \cite{Matzdorf1987} and of Aguiar et al. \cite{PhysRevC.42.2180} with the non-relativistic Rutherford scattering cross section, $d\sigma^{NR}/d\Omega$,  for $^{208}$Pb + $^{208}$Pb collisions at the laboratory energy of 100 MeV/nucleon. The dashed line is a numerical calculation following the method of  Aguiar et al. \cite{PhysRevC.42.2180}  considering relativistic corrections up to order $(v/c)^{4}$ and the solid line is a numerical calculation for the  corresponding method of Matzdorf et al. \cite{Matzdorf1987}. The horizontal axis represents the center of mass scattering angle $\Theta$ (in degrees).}
\label{fig3}
\end{figure}

\begin{figure}[ptb]
\begin{center}
\includegraphics[
width=3.4in]{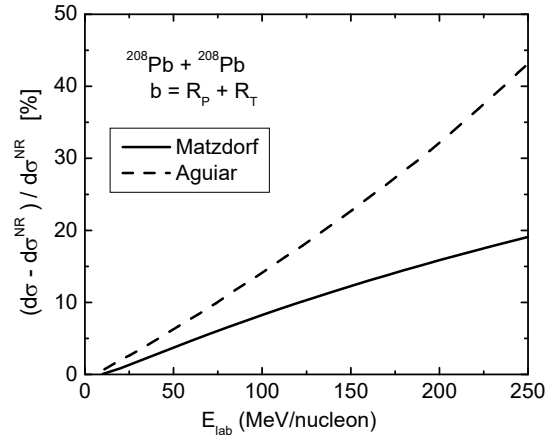}
\end{center}
\caption{Same as in Fig. \ref{fig3},  but for a collision at grazing impact parameter $b=R_{P}+R_{T}$, as function of laboratory energy $E_{lab}$ (in MeV/nucleon).}
\label{fig4}
\end{figure}

The deviations from the non-relativistic predictions are more evident for the elastic differential cross sections. This is visible in Fig. \ref{fig3} where we show the relative difference (in percent) between the numerical solutions following the methods of Matzdorf et al. \cite{Matzdorf1987} and of Aguiar et al. \cite{PhysRevC.42.2180} with the non-relativistic Rutherford scattering cross section, $d\sigma^{NR}/d\Omega$,  for $^{208}$Pb + $^{208}$Pb collisions at the laboratory energy of 100 MeV/nucleon. The dashed line follows the method of  Aguiar et al. \cite{PhysRevC.42.2180}  considering relativistic corrections up to order $(v/c)^{4}$ and the solid line is for the  corresponding method of Matzdorf et al. \cite{Matzdorf1987}. The horizontal axis represents the center of mass scattering angle $\Theta$ (in degrees). The deviations from the classical Rutherford formula clearly increase with the laboratory energy, as seen in Fig. \ref{fig4} for a collision at the grazing impact parameter. The corrections are large, almost as large as the relative change in the mass of the particles.

\begin{figure}[ptb]
\begin{center}
\includegraphics[
width=3.4in]{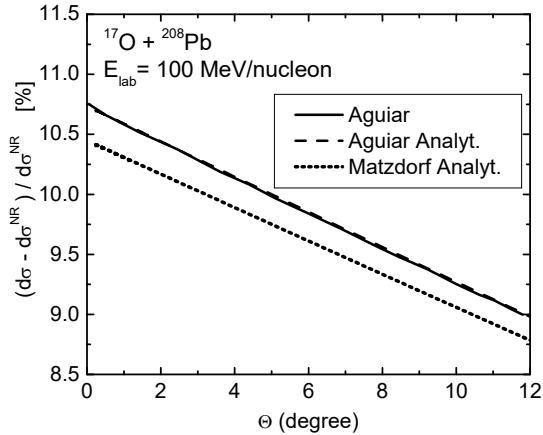}
\end{center}
\caption{Relative difference (in percent) between the analytical formulas proposed by Matzdorf et al. \cite{Matzdorf1987} and by Aguiar et al. \cite{PhysRevC.42.2180} with the non-relativistic Rutherford scattering cross section, $d\sigma^{NR}/d\Omega$,  for $^{17}$O + $^{208}$Pb collisions at the laboratory energy of 100 MeV/nucleon. The dashed line  uses the analytical equation \eqref{thea2} and the solid line is for a numerical calculation of the coupled equations following Ref. \cite{PhysRevC.42.2180}  considering relativistic corrections up to order $(v/c)^{2}$.  The dotted line is for the analytical formula \eqref{dsdo} predicted by Matzdorf et al. \cite{Matzdorf1987} which agrees within less than 0.1\% with the exact results (not shown).  }
\label{fig6}
\end{figure}

Now we turn to the precision of the analytical formulations described in the previous section that allows one to save time with numerical computations. In Fig. \ref{fig6}  we plot the relative difference (in percent) between the analytical formulas proposed by Matzdorf et al. \cite{Matzdorf1987} and by Aguiar et al. \cite{PhysRevC.42.2180} with the non-relativistic Rutherford scattering cross section, $d\sigma^{NR}/d\Omega$,  for $^{17}$O + $^{208}$Pb collisions at the laboratory energy of 100 MeV/nucleon. The dashed line is for the analytical equation \eqref{thea2} and the dashed line is for a numerical calculation following Ref. \cite{PhysRevC.42.2180}  considering relativistic corrections up to order $(v/c)^{2}$, respectively. We find that they are nearly identical. The dotted line is the analytical formula \eqref{dsdo} predicted by Matzdorf et al. \cite{Matzdorf1987} which agrees within less than 0.1\% with the exact results (not shown). Two clear conclusions from these calculations are worth mentioning: (a) the differences between the  methods of Refs. \cite{PhysRevC.42.2180} and \cite{Matzdorf1987} decrease for asymmetric systems and  (b) both analytical formulations are in excellent agreement with the corresponding models, within the range of validity of each of the two methods. The same conclusion is reached for symmetric systems.

The discussion above shows that there is no need to perform numerical calculations and solve the coupled equations proposed both in Ref. \cite{Matzdorf1987} as well as in Ref. \cite{PhysRevC.42.2180} because their proposed analytical formulations, i.e., Eqs. (\ref{theta1}) and \eqref{dsdo}, and \eqref{thea2}, yield results very close to the ``exact" numerical values. Moreover, following our numerical investigations, the method developed in Ref. \cite{Matzdorf1987} is superior than that of Ref. \cite{PhysRevC.42.2180} because it includes the full effects of retardation, which apart from the relativistic mass correction is the largest relativistic correction for the scattering of two charged particles. We have verified that the analytical formulas proposed in Ref. \cite{Matzdorf1987} both for the scattering angle and for the differential cross sections agree with numerical solutions of Eqs. \eqref{udot2}  to within 1 part in $10^3$. 

\begin{figure}[ptb]
\begin{center}
\includegraphics[
width=3.4in]{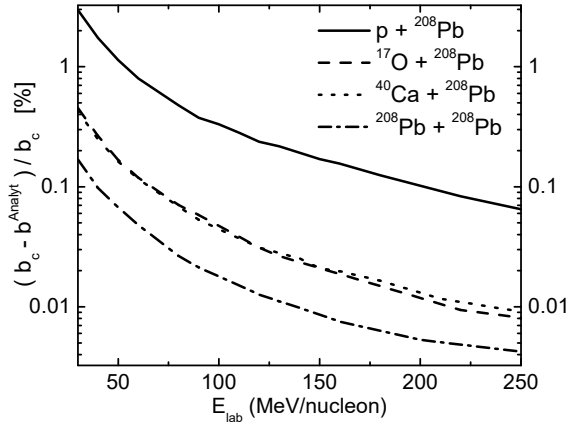}
\end{center}
\caption{Relative difference (in percent) of the distance of closest approach $b_c$ for a given impact parameter $b$ obtained with the full relativistic calculations and with  the analytical formula \eqref{bp}, as a function of laboratory energy in MeV/nucleon and for different projectile target systems. We use the grazing impact parameter $b=R_p+R_t$.}
\label{fig8}
\end{figure}

We have also studied the effects of relativity in determining the distance of closest approach between two charged particles. In Fig. \ref{fig8} we show the relative difference between  the distance of closest approach for a given impact parameter $b$ by solving Eqs. \eqref{udot1} and \eqref{udot2} and comparing it with the equation
\begin{equation}
b_c=a +\sqrt{a^2+ b^2}, \ \ \ \ \ {\rm with} \ \ \ \  a=kb={q_pq_t \over \gamma \mu v_\infty^2}, \label{bp}
\end{equation}
which is a proposed generalization of the non-relativistic relation where we replace $a_0=q_pq_t / \mu v_\infty^2$ by $a=kb=a_0/\gamma$. In the figure  we use the grazing impact parameter $b=R_p+R_t$. We see that Eq. \eqref{bp} reproduces the exact values very well at the level of 1\% or less.

Finally, we have determined the deviation of the actual time-dependent trajectory ${\bf R}(t)$ for the distance between the two charged particles from an analytical parametrization. Our parametrization is based on the same argument leading to Eq. \eqref{bp} and reads
\begin{eqnarray}
&& x = a[\cosh w +\epsilon], \ \ \ \ y=a\sqrt{\epsilon^2-1}\sinh w, \nonumber \\
&& t= {a\over v_\infty}[w +\epsilon \sinh w]. \label{par}
\end{eqnarray}
This is the same parametrization used in non-relativistic collisions \cite{Ald56}, but with the distance of closest approach $a_0$ replaced by  $a=kb=a_0/\gamma$. We have compared the difference between this approximation and the exact solution for several reaction partner combinations and energies in the range $50 - 250$ MeV/nucleon. For large times of the order of 80,000 fm we find deviations  at the level of 3\% or less. But for collision times up to  $5a/v_\infty$ after passing the distance of closest approach the Eqs.  \eqref{par} work at a much better level of 1\% or better. This explains why the distance of closest approach is so well described by the relation \eqref{bp}. This is also relevant for Coulomb excitation experiments as the Coulomb field is strongest when the trajectory is nearest to the closest approach distance, being more effective to induce nuclear transitions.

\section{Conclusions}

In this work we have studied relativistic effects such as retardation, relativistic mass change, and the inclusion of magnetic interactions in the Coulomb scattering of nuclei at intermediate and high energies ($E_{lab} \gtrsim 50$ MeV/nucleon). Several conclusions have been drawn from this work. We have shown that the formalism developed in Ref.  \cite{Matzdorf1987} provides a concise way to obtain Coulomb scattering deflection angles and elastic differential cross sections. Their method is superior than the one proposed in Ref. \cite{PhysRevC.42.2180} with an effective Lagrangian expansion in orders of $v/c$. 

Most importantly, we have found that analytical equations are able to describe the exact results obtained with the numerical solutions of Eqs. \eqref{udot2}. The deflection angle is well described by Eq. \eqref{theta1} while the differential cross section is well described by Eq.  \eqref{dsdo}. Finally, the distance of closest approach for a given impact parameter $b$, as well as the time dependence of the trajectory are in good agreement with the Eqs. \eqref{bp} and \eqref{par}, respectively.  

These findings are timely and of importance for the experimental analysis of numerous data being acquired in radioactive beam facilities with laboratory energies in the range of $E_{lab} \gtrsim 50$ MeV/nucleon. The determination of Coulomb scattering angles and differential cross sections are a crucial part of the simulations and  the extraction of reaction variables.

\section*{Acknowledgement} 

This work was supported in part by the U.S. DOE grants DE-FG02-08ER41533 and  the U.S. NSF Grant No. 1415656.   Ravinder Kumar is Raman Postdoctoral Fellow during 2016-17 through the University Grants Commission, New Delhi, India, Grant No. F. No. 5-152/2016(IC).

%\bibliography{sofftext}{}
%\bibliographystyle{unsrt}

%\bibliographystyle{apsrev}

\end{document}